\documentclass[letterpaper, 10 pt, conference]{ieeeconf}  

\IEEEoverridecommandlockouts                              

\usepackage{epsfig}
\usepackage{amsmath}
\usepackage{cleveref}
\usepackage{footmisc}
\usepackage{color}
\usepackage{xcolor}
\usepackage{mathrsfs}
\usepackage{multirow}
\usepackage{textcomp}
\usepackage{amssymb}
\usepackage{array}
\definecolor{qb}{rgb}{0.9, 0.17, 0.31}
\title{Instantiation-Net: 3D Mesh Reconstruction from Single 2D Image for Right Ventricle}

\author{Zhao-Yang Wang$^{1*}$ and Xiao-Yun Zhou$^{1*}$ and Peichao Li$^{1}$ and Celia Riga$^{3}$ and Guang-Zhong Yang$^{1,2}$
\thanks{*Zhao-Yang Wang and Xiao-Yun Zhou contribute equally to this paper}
\thanks{$^{1}$The Hamlyn Centre for Robotic Surgery, Imperial College London, UK
        {\tt\small wangzybmie@outlook.com, xiaoyun.zhou14@imperial.ac.uk}}%
\thanks{$^{2}$Institute of Medical Robotics, Shanghai Jiao Tong University, China}%
\thanks{$^{3}$Regional Vascular Unit, St Mary’s Hospital, London, UK and the Academic Division of Surgery, Imperial College London, UK}%
}

\begin{document}

\maketitle
\thispagestyle{empty}
\pagestyle{empty}

\begin{abstract}

3D shape instantiation which reconstructs the 3D shape of a target from limited 2D images or projections is an emerging technique for surgical intervention. It improves the currently less-informative and insufficient 2D navigation schemes for robot-assisted Minimally Invasive Surgery (MIS) to 3D navigation. Previously, a general and registration-free framework was proposed for 3D shape instantiation based on Kernel Partial Least Square Regression (KPLSR), requiring manually segmented anatomical structures as the pre-requisite. Two hyper-parameters including the Gaussian width and component number also need to be carefully adjusted. Deep Convolutional Neural Network (DCNN) based framework has also been proposed to reconstruct a 3D point cloud from a single 2D image, with end-to-end and fully automatic learning. In this paper, an Instantiation-Net is proposed to reconstruct the 3D mesh of a target from its a single 2D image, by using DCNN to extract features from the 2D image and Graph Convolutional Network (GCN) to reconstruct the 3D mesh, and using Fully Connected (FC) layers to connect the DCNN to GCN. Detailed validation was performed to demonstrate the practical strength of the method and its potential clinical use.

\end{abstract}

\section{Introduction}
\label{sec:intro}

Recent advances in robot-assisted Minimally Invasive Surgery (MIS) brings many advantages to patients including reduced access trauma, less bleeding and shorter hospitalisation. They also bring challenges in terms of articulated instrument design and advanced intra-operative visualisation and guidance for targetting small, deep seated lesions. For pre-operative planning, 3D imaging techniques including Magnetic Resonance Imaging (MRI), Computed Tomography (CT), and ultrasound can be used. For intra-operative guidance, acquisition of 3D data in real-time is challenging and in most clinical practices, 2D projections or images from fluoroscopy, MRI and ultrasound are used only.
It is difficult to use these 2D images to resolve complex 3D geometries and therefore there is a pressing need to develop computer vision techniques to reconstruct 3D structures from a limited number of 2D projections or images in real-time intra-operatively.


\begin{figure}
    \centering
    \framebox{\includegraphics[width=0.4\textwidth]{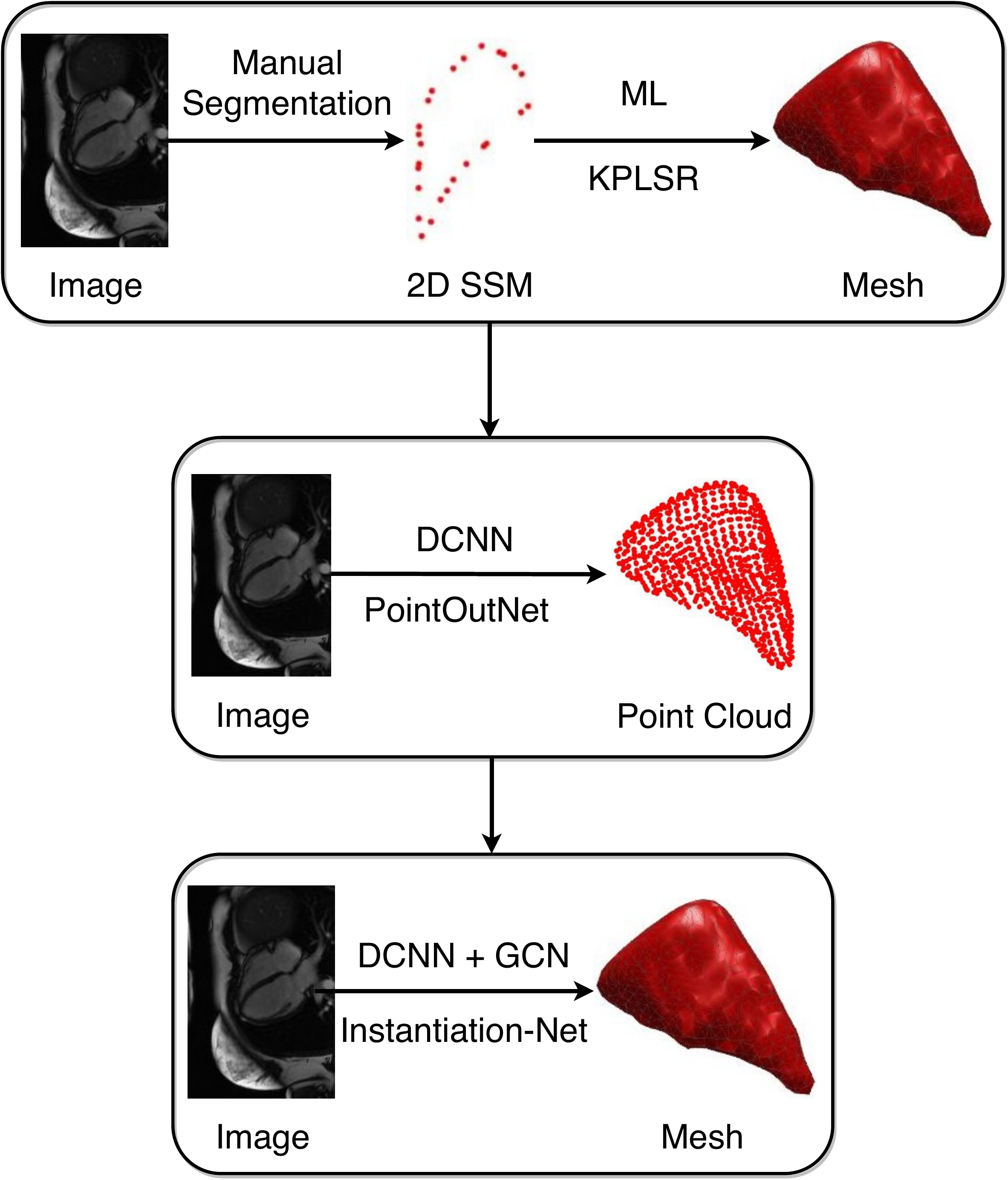}}
    \caption{An illustration of the evolution of the framework in 3D shape instantiation from two-stage approach based on KPLSR (top) \cite{zhou2018real}, point cloud output by PointOutNet (middle) \cite{zhou2019one} to the Instantiation-Net (bottom) that could reconstruct 3D mesh from 2D image in end-to-end fashion.}
    \label{fig:intro}
\end{figure}

A traditional and popular framework is to reconstruct the 3D shape of a target pre-operatively from CT or MRI volume scans, then register this 3D shape to intra-operative 2D projections. However, registration based methods are computationally demanding and static. In practice, due to the influence of respiration and patient's motion, the operation area is usually deformed where dynamic and real-time 3D navigation is essential. Recently, 3D shape instantiation is emerging which reconstructs the 3D shape of a target from limited or a single 2D projection or image of the target in real-time. It usually has two main requirements: 1) low computational complexity; 2) limited input images.

For example, the 3D shape of a stent-graft at three different statuses: fully-compressed \cite{zhoustent}, partially-deployed \cite{zheng2019real} and fully-deployed \cite{zhou2018real-instantiation} was instantiated from a single 2D fluoroscopic projection with stent-graft modelling, graft gap interpolation, the Robust Perspective-n-Point (RPnP) method, Graph Convolutional Network (GCN), and mesh manipulation. The 3D Abdominal Aortic Aneurysm (AAA) skeleton was instantiated from a single intra-operative 2D fluoroscopic projection with graph matching and skeleton deformation \cite{zheng2019towards}.

Recently, a general and registration-free framework for 3D shape instantiation was proposed in \cite{zhou2018real} with three steps: 1) multiple 3D volumes were pre-operatively scanned for the target at different time frames during the deformation cycle. 3D meshes were segmented manually from the volumes and were expressed into 3D Statistical Shape Model (SSM) where all meshes were expressed with the same number of vertex and the same connectivity. Sparse Principal Component Analysis (SPCA) was used to analyse the 3D SSM to determine the most informative and important scan plane; 2) multiple 2D images were scanned synchronously to the 3D volumes at the determined optimal scan plane. 2D contours were segmented manually from the images and were expressed into 2D SSM. Kernel Partial Least Square Regression (KPLSR) was applied to learn the relationship between the 2D SSM and 3D SSM; 3) the KPLSR-learned relationship was applied on the intra-operative 2D SSM to reconstruct the instantaneous 3D SSM for navigation. Two deficiencies exist in \cite{zhou2018real}: 1) during the intra-operative application, 2D SSM needs to be generated from the intra-operative 2D image and then to be used as the input for the KPLSR-based mesh reconstruction; 2) two hyper-parameters including the Gaussian width and component number require to be carefully and manually adjusted.

To avoid these drawbacks, a one-stage and fully automatic framework for 3D shape instantiation was proposed to reconstruct the 3D point cloud of a target from its a single 2D projection with PointOutNet and Chamfer loss in \cite{zhou2019one}. Although it achieved end-to-end learning from 2D image to 3D point cloud, the vertex correspondence lost due to the compulsory involvement of Chamfer loss, and therefore 3D mesh can not be reconstructed.

Compared to a two-stage method, directly reconstructing 3D shape from one single 2D image will have boarder and more practical applications in both natural and medical computer vision. In \cite{choy20163d}, based on Long Short-Term Memory (LSTM), a 3D Recurrent Reconstruction Neural Network (3D-R2N2) was proposed to recover 3D volume from one single real-world image. However, 3D volume is low resolution. Compared to un-ordered 3D point cloud or low-resolution 3D volume, 3D mesh with more details of the surface is more helpful and vital for surgical 3D navigation. Based on the CoMA auto-encoder in \cite{ranjan2018generating}, \cite{kulon2019single} proposed a two-stage method to recover 3D hand mesh from single 2D image. In \cite{wang2018pixel2mesh}, an end-to-end graph based convolutional neural network was proposed to recover 3D mesh from single 2D real-world image. However, the proposed method required an initial input of an ellipsoid mesh together with the image perceptual feature, which was then gradually deformed into the target mesh.     

In this paper, we propose the Instantiation-Net to reconstruct the 3D mesh of a target from its a single 2D projection. Deep Convolutional Neural Network (DCNN) - DenseNet-121 with convolutional layers and average pooling layers is used to extract abundant features from the 2D image input. Graph Convolutional Network (GCN) is used to reconstruct the 3D mesh including the vertex coordinates and connectivity. Fully Connected (FC) layers are used to pass the features extracted from the DCNN to be as the input for the GCN. The Fig. \ref{fig:intro} illustrates the framework for 3D shape instantiation is evolving from two-stage KPLSR-based framework \cite{zhou2018real}, the PointOutNet \cite{zhou2019one}, to the Instantiation-Net proposed in this paper.

Following \cite{zhou2018real, zhou2019one}, 27 Right Ventricles (RVs) were scanned at multiple different time frames of the cardiac cycle with 3D volumes and synchronized 2D images. Patient-specific training was used with the leave-one-out cross validation, indicating 609 experiments. An average 3D distance error around $2mm$ was achieved, which is comparable to the performance in \cite{zhou2018real} but with end-to-end and fully automatic training. Details of the proposed Instantiation-Net including the DCNN, GCN and FC part and experimental setup are stated in Sec. \ref{sec:method}. The results with validating the proposed Instantiation-Net on 27 RVs are illustrated in Sec. \ref{sec:result}, followed by the discussion in Sec. \ref{sec:discussion} and the conclusion in Sec. \ref{sec:conclusion}.

\section{Methodology}
\label{sec:method}

The input of Instantiation-Net is a single 2D image $I$ with a size of $192 \times 256$ while the output is a 3D mesh $\mathscr{F}$ with vertex $V$ and the connectivity $A$,

\subsection{DCNN}
\label{sec:method-dcnn}

For an image input $I_{\rm N\times \rm H\times \rm W \times \rm C}$, where $\rm N$ is the batch size and is fixed at 1 in this paper, $\rm H$ is the image height, $\rm W$ is the image width, $\rm C$ is the image channel and is 1 for medical images, convolutional and pooling layers are applied to extract feature maps $F_{\rm N\times \rm H\times \rm W \times \rm C}$.

In convolutional layers, a trainable convolutional kernel $T_{\rm C_{in} \times K \times K \times C_{out}}$, where $\rm K$ is the kernel size, moves along the height and width of the feature map with a stride of S:

\begin{equation}
\hat{F}_{\rm N \times H' \times W' \times C_{out}} = F_{\rm N \times H \times W \times C_{in}} \cdot  T_{\rm C_{in} \times K \times K \times C_{out}}
\end{equation}

where $\rm H'=H//S$, $\rm W'=W//S$, $//$ is floor division. In pooling layers, a kernel $T_{\rm K \times K}$ moves along the height and width of a feature map with a stride of $\rm K$:

\begin{equation}
\hat{F}_{\rm N \times H' \times W' \times C} = F_{\rm N \times H \times W \times C} \cdot  T_{\rm K \times K}
\end{equation}

where $\rm H'=H//K$, $\rm W'=W//K$. The maximum or average value of the $\rm K \times K$ receptive field is selected to represent this area, indicating max-pooling or average-pooling layers.

Batch Normalization (BN) layers are applied to facilitate the DCNN training, with the mean and variance calculated as:

\begin{equation}
\mu_c = \frac{1}{\rm N \times \rm H \times \rm W} \sum\limits_{n=1}^{\rm N} \sum\limits_{h=1}^{\rm H} \sum\limits_{w=1}^{\rm W} f_{n,h,w}
\end{equation}

\begin{equation}
\delta_c^2 = \frac{1}{\rm N \times \rm H \times \rm W} \sum\limits_{n=1}^{\rm N} \sum\limits_{h=1}^{\rm H} \sum\limits_{w=1}^{\rm W} (f_{n,h,w}-\mu_c)^2
\end{equation}

Then $F$ is normalized to be with a mean of 0 and a variance of 1, re-scaled by $\gamma_c$ and re-translated by $\beta_c$ to maintain the DCNN representation ability as:

\begin{equation}
\hat{f}_{n,h,w} = \frac{f_{n,h,w}-\mu_c}{\sqrt{\delta_c^2+\epsilon}} \gamma_c + \beta_c
\end{equation}

where $\epsilon$ is a small value which is added for division stability. In addition, Rectified Linear Unit (ReLU) layers $f(x) = max (0, x)$ are also used to add non-linearity.

Multiple convolutional layers, BN layers, average-pooling layers and ReLU layers consist the first part of the proposed Instantiation-Net in this paper - DenseNet-121 \cite{huang2017densely} for extracting abundant features from the single 2D image input.

\subsection{GCN}
\label{sec:method-gcn}

For a 3D mesh $\mathscr{F}$ with vertex of $\mathbf{V}_{\rm M \times 3}$ and connectivity of $\mathbf{A}_{\rm M \times M}$, where $\rm M$ is the number of vertex in the mesh, $\mathbf{A}_{\rm M \times M}$ is the adjacency matrix with, $\mathbf{A}_{ij} = 1$ if the $ith$ and $jth$ vertex are connected by an edge, otherwise $\mathbf{A}_{ij} = 0$. The non-normalized graph Laplacian matrix is calculated as:

\begin{equation}
    \rm \mathbf{L}=\mathbf{D}-\mathbf{A}
\end{equation}

where $\mathbf{D}_{ii} = \sum_{j=1}^{\rm M} \mathbf{A}_{ij}$, $\mathbf{D}_{ij}=0$, if $i\neq j$ is the vertex degree matrix.

For achieving Fourier transform on the mesh vertex, $L$ is decomposed into Fourier basis as:

\begin{equation}
\mathbf{L}=\mathbf{U} \Lambda \mathbf{U}^T
\end{equation}

where $\mathbf{U}$ is the matrix of eigen-vectors and $\Lambda$ is the matrix of eigen-values. The Fourier transform on the vertex $v$ is then formulated as $v_w = U^Tv$, while the inverse Fourier transform is formulated as $v = \mathbf{U}^Tv_w$. The convolution in spatial domain of the vertex $v$ and the kernel $s$ can be inversely transformed from the spectral domain as:

\begin{equation}
v*s = \mathbf{U}((\mathbf{U}^Tv)\odot(\mathbf{U}^Ts))
\end{equation}

where $s$ is the convolutional filter. However, this computation is very expensive as it involves matrix multiplication. Hence Chebyshev polynomial is used to reformulate the computation with a kernel $g_{\theta}$:

\begin{equation}
    g_{\theta}(\mathbf{L})=\sum_{k=0}^{K-1}\theta_{k}T_{k}(\tilde{\mathbf{L}})
\end{equation}

where $\tilde{\mathbf{L}} = 2\mathbf{L}/\lambda_{max}-\mathbf{I}_n$ is the scaled Laplacian, $\theta$ is the Chebyshev coefficient. $T_{k}$ is the Chebyshev polynomial and is recursively calculated as~\cite{defferrard2016}:

\begin{equation}
    \mathbf{T}_{k}(\tilde{\mathbf{L}}) = 2\tilde{\mathbf{L}}\mathbf{T}_{k-1}(\tilde{\mathbf{L}})-\mathbf{T}_{k-2}(\tilde{\mathbf{L}})
\end{equation}

where $\mathbf{T}_0=1$, $\mathbf{T}_1=\tilde{\mathbf{L}}$. The spectral convolution is then defined as:

\begin{equation}
    y_j = v*s = \sum_{i=1}^{F_{in}}g_{\theta i,j}(\mathbf{L})v
\end{equation}

where $F_{in}$ is the feature channel number of the input $V$, $j\in (1, F_{out})$, $F_{out}$ is the feature channel number of the output $Y$. Each convolutional layer has $F_{in}\times F_{out} \times K$ trainable parameters.

Except graph convolutional layers, up-sampling layers are also applied to learn the hierarchy mesh structures. First, the mesh $\mathscr{F}$ is down-sampled or simplified to a simplified mesh with $\rm M//S$ vertices, where $\rm S$ is the stride, and is set as 4 or 3 in this paper. Several mesh simplification algorithms can used in this stage, such as Quadric Error Metrics \cite{ranjan2018generating, garland1997surface}, and weighted Quadric Error Metrics Simplification (QEMS) \cite{zhou2016path}, The connectivity of the simplified meshes are recorded and used to calculate $\mathbf{L}$ for the graph convolution at different resolutions. The discarded vertexes during the mesh simplification are projected back to the nearest triangle, with the projected position computed with the barycentric coordinates. More details regarding the down-sampling, up-sampling and graph convolutional layers can be found in \cite{ranjan2018generating}.

\begin{table}
\label{tab:network}
\centering
\caption{Detailed layer configurations of the proposed Instantiation-Net including the DCNN, FC and GCN part.}
\begin{tabular}{|l|l|}

\multicolumn{2}{c}{ }\\
\multicolumn{2}{c}{\textbf{DCNN}} \\
\multicolumn{2}{c}{ }\\ \hline

\hline
\textbf{Layers}     & \textbf{Configuration}  \\ \hline
Convolution         & 7$\times$ 7 conv, stride 2 \\ \hline
Pooling             & 3$\times$ 3 max pool, stride 2 \\ \hline
Dense Block(1)      & $\left[ \begin{array}{ccc}
	                               1 \times 1 &conv\\
	                               3 \times 3 &conv\\
	                               \end{array} \right]  \times 6$, stride 1 \\ \hline
\multirow{2}{*}{Transition Layer(1)}&1 $\times$ 1 conv \\ & 2 $\times 2$ average pool, stride 2 \\\hline 

Dense Block(2)     & $\left[ \begin{array}{ccc}
	                               1 \times 1 &conv\\
	                               3 \times 3 &conv\\
	                               \end{array} \right]  \times 12$, stride 1       \\ \hline
\multirow{2}{*}{Transition Layer(2)} &1$\times$1 conv \\& 2 $\times$ 2 average pool, stride 2\\\hline 

Dense Block(3)     & $\left[ \begin{array}{ccc}
	                               1 \times 1 &conv\\
	                               3 \times 3 &conv\\
	                               \end{array} \right]  \times 24$, stride 1      \\ \hline 
\multirow{2}{*}{Transition Layer(3)} &1$\times$1 conv \\& 2 $\times$ 2 average pool, stride 2 \\\hline

Dense Block(4)     & $\left[ \begin{array}{ccc}
	                               1 \times 1 &conv\\
	                               3 \times 3 &conv\\
	                               \end{array} \right]  \times 16$, stride 1       \\ \hline
\multicolumn{2}{c}{ }\\
\multicolumn{2}{c}{\textbf{FC}} \\
\multicolumn{2}{c}{ }\\ \hline

\hline
\textbf{Layers}     & \textbf{Configuration}  \\ \hline
Fully Connected     &            $[1, 6, 8, 1024] \xrightarrow{} 8$             \\ \hline
Fully Connected     &            $8 \xrightarrow{} (M / 4^{4}) \times 64$              \\ \hline 

\multicolumn{2}{c}{ }\\
\multicolumn{2}{c}{\textbf{GCN}} \\
\multicolumn{2}{c}{ }\\ \hline

\hline
\textbf{Layers}     & \textbf{Configuration}  \\ \hline
Up-Sampling         & $(M / 4^{4}) \times 64 \xrightarrow{}(M / 4^{3}) \times 64$      \\ \hline
Convolution         & $(M / 4^{3}) \times 64 \xrightarrow{}(M / 4^{3}) \times 64$      \\ \hline
Up-Sampling         & $(M / 4^{3}) \times 64 \xrightarrow{}(M / 4^{2}) \times 64$      \\ \hline
Convolution         & $(M / 4^{2}) \times 64 \xrightarrow{}(M / 4^{2}) \times 64$      \\ \hline
Up-Sampling         & $(M / 4^{2}) \times 64 \xrightarrow{}(M / 4) \times 64$          \\ \hline
Convolution         & $(M / 4) \times 64     \xrightarrow{}(M / 4) \times 64$          \\ \hline
Up-Sampling         & $(M / 4) \times 64     \xrightarrow{}M \times 64$                \\ \hline
Convolution         &  $M \times 64          \xrightarrow{}M \times 3$                \\ \hline
\end{tabular}
\end{table}

\begin{figure*}
    \centering
    \framebox{\includegraphics[width=0.97\textwidth]{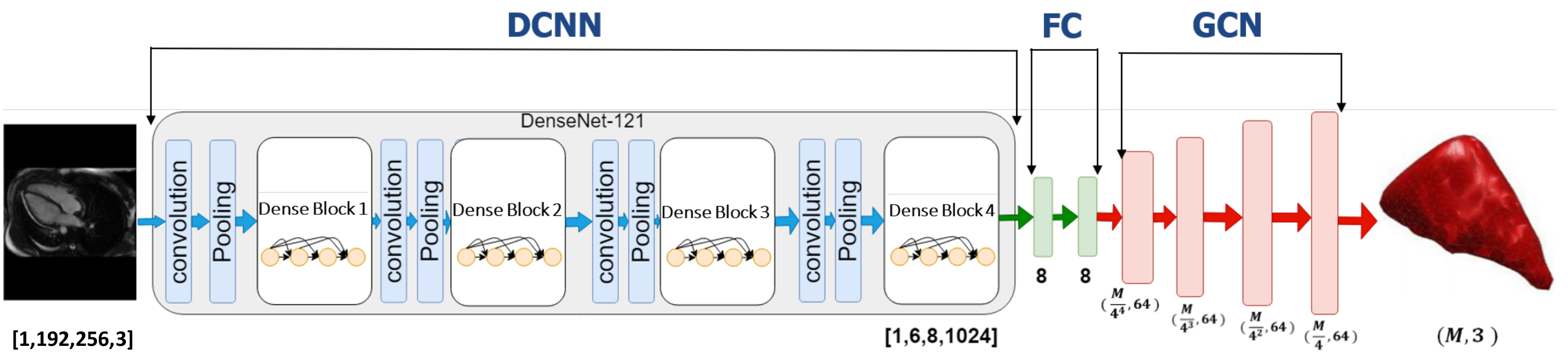}}
    \caption{An intuitive illustration of the proposed Instantiation-Net with compacting multiple layers into blocks, with three parts DCNN, FC and GCN.}
    \label{fig:network}
\end{figure*}

\subsection{Instantiation-Net}
\label{sec:method-instantiation}

For the DCNN part, DenseNet-121 from \cite{huang2017densely} is imported from Keras, with parameters pre-trained on the ImageNet \cite{deng2009imagenet}. For the FC part, two FC layers with an output feature dimension of 8 are used. For the GCN part, four up-sampling and graph convolutional layers are adopted from \cite{ranjan2018generating}. An example of the proposed Instantiation-Net including the DCNN, FC and GCN part is shown in Tab. \ref{tab:network} with detailed configurations of each layer. Convolutional layers, max-pooling layers, Dense blocks and transition layers are applied in the DCNN part. Two FC layers are used in the FC part. Four graph convolutional layers and up-sampling layers are used in the GCN part. An intuitive illustration of the example Instantiation-Net with compacting multiple layers into blocks is also shown in Fig.~\ref{fig:network}. The input is generated by tiling the 2D MRI image three times along the channel dimension. A 3D mesh can be reconstructed directly from the single 2D image input by the proposed Instantiation-Net in a fully automatic and end-to-end learning fashion.

\subsection{Experimental Setup}
\label{sec:method-experiment}

\subsubsection{Data collection}

27 RVs including 18 asymptomatic and 9 Hypertrophic Cardiomyopathy (HCM) subjects were scanned by a 1.5T MRI scanner (Sonata, Siemens, Erlangen, Germany), with a pixel spacing of $1.5-2mm$. For each patient, the cardiac deformation cycle was divided into $19-25$ time frames. In the training stage, the 2D images and 3D meshes are acquired for each patient at each time frame, indicating 609 pairs of training data.

In terms of the acquisition of the 2D images of the RV for training, a single 2D MRI image was scanned at the long-axis with viewing the four chambers. The scan time frame was synchronized to the scanning time frame of the 3D volumes with the same MRI machine. One 3D mesh and one 2D image consisted of one pair of training data. This data was also widely used as the validation for 3D shape instantiation task in \cite{zhou2018real, zhou2019one}, and 2D medical image segmentation task in \cite{zhou2019normalization, zhou2019u, zhou2019atrous}.

In order to obtain the 3D meshes of the RV for training, ten 2D MRI images were scanned from the atrioventricular ring to the apex with $10mm$ slice gap. Analyze (AnalyzeDirect, Inc, Overland Park, KS, USA) was used to interpolate ten new slices between each two images. It was also used to segment the RV contours manually from interpolated slices. Meshlab \cite{Meshlab} was used to reconstruct the 3D meshes from the segmented 2D RV contours and to smooth the reconstructed 3D meshes with smoothing filters. For each patient, 3D meshes of the RV are generated as 3D SSM which is a format of mesh with consistent number of vertex and consistent connectivity across the different time frames between 19 to 25. To note that, each mesh across different time frames is synchronized with the 2D images.

\subsubsection{Training configuration}
Following \cite{zhou2018real, zhou2019one}, the proposed Instantiation-Net was trained patient-specifically with leave-one-out cross-validation: one time frame in the patient was used in the test set,  while all other time frames were used as the training set. Stochastic Gradient Descent (SGD) was used as the optimizer with a momentum of 0.9, while each experiment was trained up to 1200 epochs. The initial learning rate was $5e^{-3}$ and decayed with 0.97 every $5\times \rm M$ iterations, where $\rm M$ is the number of time frame for each patient. The kernel size of GCN was 3. For most experiments, the feature channel and stride size of GCN were 64 and 4 respectively, except that there some experiments used 16 and 3 instead. The proposed framework was implemented on Tensorflow and Keras functions. L1 loss was used as the loss function, because L2 loss experienced convergence difficulty in our experiments:

\begin{equation}
    Loss_{L1} =\frac{1}{\rm 3M } \sum_{m=1}^{\rm 3M}|g_m - p_m|
\end{equation}

where $p_m$ is one vertex coordinate predicted by the proposed Instantiation-Net, while $g_m$ is the corresponding vertex coordinate in the ground truth. The 3D distance error in Eq.~\ref{eq:3D_dist} is the average value of the 3D distance errors of all the vertices in the 3D mesh. It is used as the evaluation metric:

\begin{equation}\label{eq:3D_dist}
    E =\frac{1}{\rm M} \sum_{m=1}^{\rm M} \sqrt {\sum_{i=1}^3(g_{m,i} - p_{m,i})^2}
\end{equation}

where, $g$ and $p$ are the reformulated ground truth and prediction with a size of $(\rm M, 3)$.

\section{Result}
\label{sec:result}
To prove the stability and robustness of the proposed Instantiation-Net to each vertex inside a mesh and to each time frame inside a patient, the 3D distance error for each vertex of four meshes and 3D distance error for each time frame of 12 patients are shown in Sec. \ref{sec:result-mesh} and Sec. \ref{sec:result-patient} respectively. To validate the performance of the proposed Instantiation-Net, the PLSR-based and KPLSR-based 3D shape instantiation in \cite{zhou2018real} are adopted as the baseline with validating on the same 27 subjects in Sec. \ref{sec:result-method}

\subsection{3D distance error for a mesh}
\label{sec:result-mesh}

\begin{figure}
    \centering
    \framebox{\includegraphics[width=0.4\textwidth]{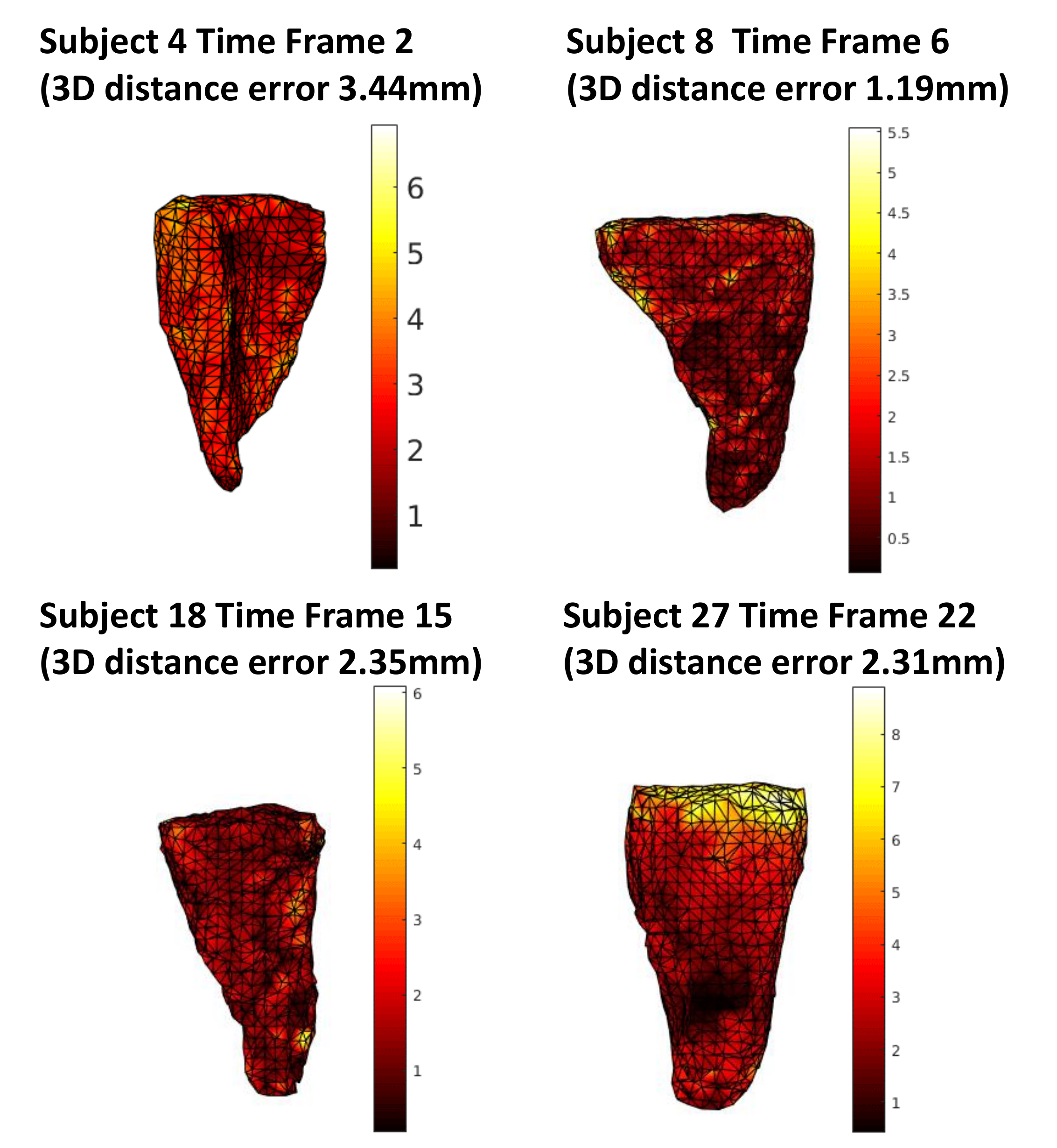}}
    \caption{The 3D distance error of each vertex of four randomly-selected reconstructed meshes, with the brighter color indicating the higher error value. The color bar is in a unit of $mm$.}
    \label{fig:mesherror}
\end{figure}

Four reconstructed meshes were selected randomly, with showing the 3D distance error of each vertex in colors in Fig. \ref{fig:mesherror}. We can see that the error is distributed equally on each vertex, as the color on the mesh is very similar. The error does not concentrate or cluster on one specific area. High errors appear at the top of the RV, which is normal, as the vertex number at the RV mesh top is much less in the ground truth than other areas. This will be further discussed in Sec. \ref{sec:discussion}.

\subsection{3D distance error for a patient}
\label{sec:result-patient}

\begin{figure}
    \centering
    \framebox{\includegraphics[width=0.47\textwidth]{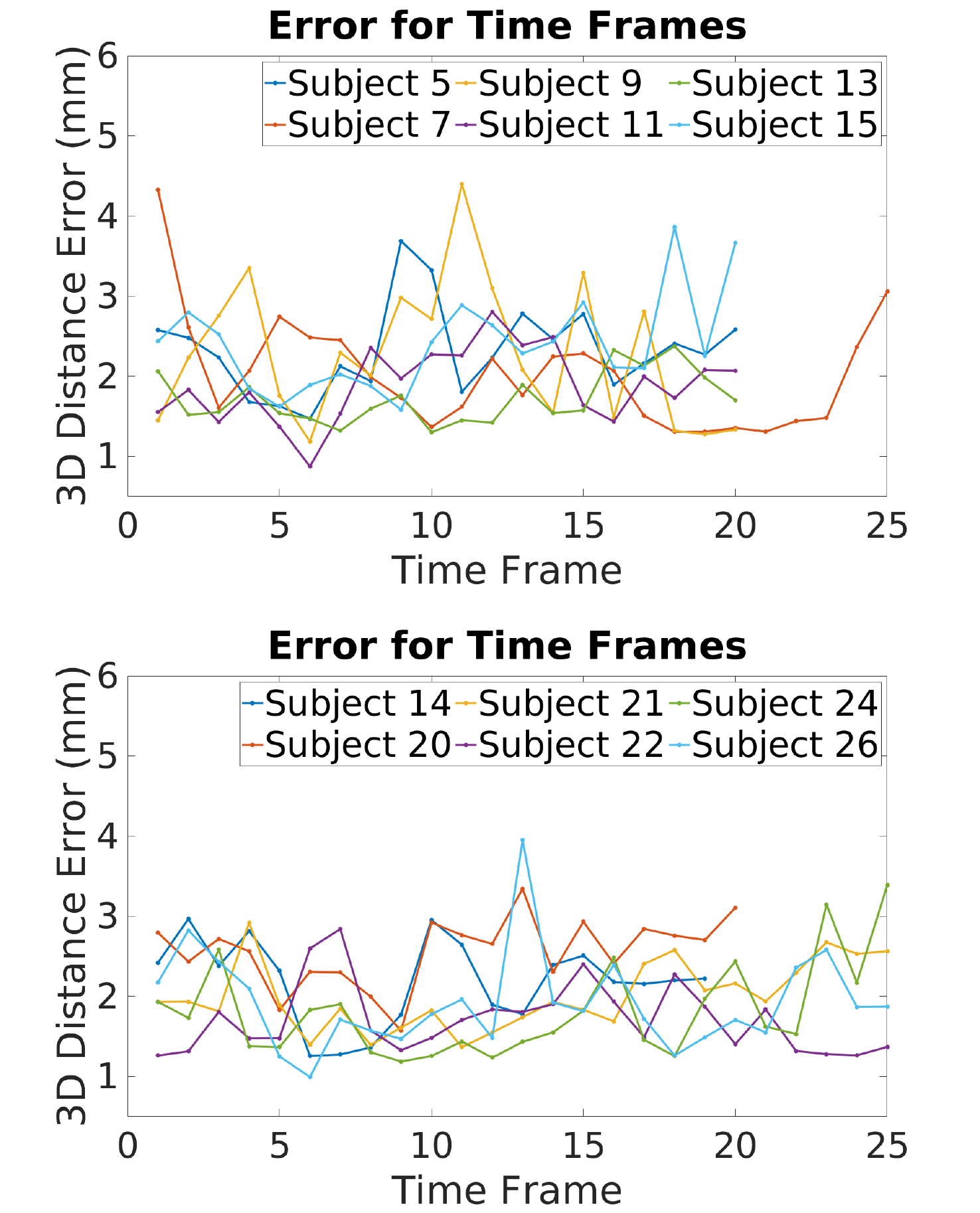}}
    \caption{The mean 3D distance errors of each time frame of 12 randomly selected patients.}
    \label{fig:patienterror}
\end{figure}

The Fig. \ref{fig:patienterror} illustrates the 3D distance errors of each time frame of 12 subjects selected randomly from 27 subjects. Note that, the 3D distance error of each time frame is calculated by Eq.~\ref{eq:3D_dist}. We can see that, for most time frames, the 3D distance errors are around $2mm$. High errors appear at some time frames, i.e. the time frame 1 and 25 of subject 7, the time frame 11 of subject 9, the time frame 9 of subject 5, the time frame 18 and 20 of subject 15, the time frame 13 of subject 26. This phenomenon was also observed in \cite{zhou2018real, zhou2019one}, which is due to boundary effect. At the systole or diastole moment of the cardiac, the shape of cardiac reaches its smallest or largest size, resulting in extreme cases of 3D mesh compared with other time frames. In leave-one-out cross validation, if these extreme time frames are not seen in the training data, but are used in the inference stage of the network, the accuracy of the prediction will be lower. This will also be further discussed in Sec. \ref{sec:discussion}.

\subsection{Comparison to other methods}
\label{sec:result-method}
\begin{figure*}
    \centering
    \framebox{\includegraphics[width=0.6\textwidth]{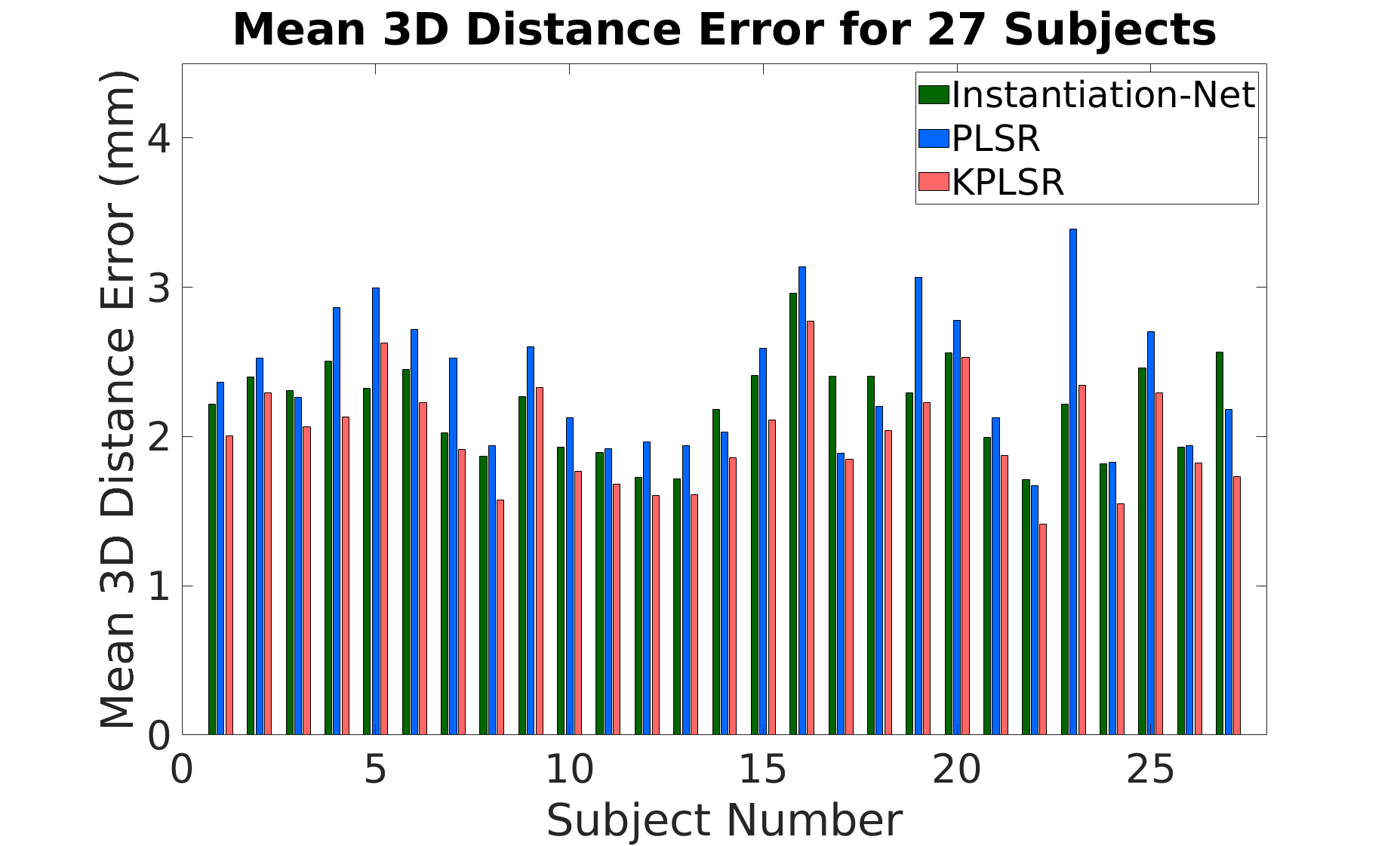}}
    \caption{The mean 3D distance errors of the mesh of 27 subjects generated by the proposed Instantiation-Net, PLSR- and KPLSR-based 3D shape instantiation.}
    \label{fig:overallerror}
\end{figure*}

The Fig. \ref{fig:overallerror} shows the comparison of the performance among the proposed Instantiation-Net, PLSR- and KPLSR-based 3D shape instantiation on the 27 subjects. These were evaluated by the mean of 3D distance errors $E$ across all time frames. We can see that the proposed Instantiation-Net out-performs PLSR-based 3D shape instantiation while under-performs KPLSR-based 3D shape instantiation slightly for most patients. The overall mean 3D distance error of the mesh generated by the proposed Instantiation-Net, PLSR-based and KPLSR-based 3D shape instantiation are $2.21mm$, $2.38mm$ and $2.01mm$ respectively. In addition, the performance of the proposed Instantiation-Net is robust across different patients, as no obvious outliers are observed among the 27 subjects.

All experiments were trained with a CPU of Intel Xeon\textsuperscript{\textregistered} E5-1650 v4@3.60GHz$\times$12 and a GPU of Nvidia Titan Xp. The GPU memory consuming was around 11G which is larger than the PointOutNet in \cite{zhou2019one} which consumed 4G, while PLSR-based and KPLSR-based method in \cite{zhou2018real} were trained on a CPU. The training time was around 1h for one time frame which is longer than that of the PointOutNet in \cite{zhou2019one} which was around $30mins$, while PLSR-based and KPLSR-based method in \cite{zhou2018real} took a few minutes. However, the inference of the end-to-end Instantiation-Net only took 0.5 seconds to generate a 3D mesh automatically, without the need of the manual segmentation in KPLSR-based 3D shape instantiation.

\section{Discussion}
\label{sec:discussion}

During the process of segmenting the RV contour from the 2D MRI short-axis image,  only $1-2$ images could be  scanned at the atrioventricular ring, due to the sudden termination of this anatomy. Hence less vertexes and sparse connectivity exist at the top of the 3D RV mesh, resulting in a higher error in this area, as shown in the right-bottom example in Fig. \ref{fig:mesherror}. This phenomenon also happened in \cite{zhou2018real, zhou2019one}.

During the cardiac beating, the RV deforms into the smallest and largest size in the systole and diastole time frame respectively. In the leave-one-out cross validation, if either of the systole and diastole time frame are not seen in the training set during training but used in the testing set, higher 3D distance error occurs. This phenomenon is called boundary effect and also occurred in \cite{zhou2018real, zhou2019one}, i.e. the 1st and 25th time frame of subject 7 (Fig.~\ref{fig:patienterror}). In practical applications, the training data will cover all time frames pre-operatively, which eliminates this phenomenon.

In \cite{zhou2018real}, PLSR-based and KPLSR-based 3D shape instantiation were proposed. However, manual segmentation was needed to generate the 2D SSM as the model input while manual tuning on the two hyper-parameters were vital for a good performance. In order to achieve fully automatic and end-to-end learning, PointOutNet was proposed in \cite{zhou2019one} to be trained along with the Chamfer loss which showed better convergence ability than the traditional L1 and L2 loss. However, the vertex correspondence was lost and hence only 3D point cloud rather than 3D mesh was achieved. For achieving 3D mesh reconstruction, the Instantiation-Net is proposed in this paper to reconstruct the 3D mesh of a target from its a single 2D image in a fully automatic and end-to-end training fashion.

DCNN has a powerful ability for features extraction from images while GCN has a powerful ability for mesh deformation with both vertex deformation and connectivity maintenance. This paper integrates these two strong networks to achieve 3D mesh reconstruction from 2D image, which crosses modalities. Based on the author's knowledge, this is one of the few pioneering works that achieve direct 3D mesh reconstruction from 2D images. In medical computer vision, this is the first work that achieves 3D mesh reconstruction from a single 2D image in an end-to-end and fully automatic training fashion.

One potential shortage of the proposed Instantiation-Net is that it requires both the larger consumption in GPU memory and the longer training time than that of the PointOutNet in \cite{zhou2018real} and the PLSR-based and KPLSR-based 3D shape instantiation in \cite{zhou2019one}.

\section{Conclusion}
\label{sec:conclusion}
In this paper, an end-to-end framework, called Instantiation-Net, was proposed to instantiate the 3D mesh of RV from its a single 2D MRI image. DCNN is used to extract the feature map from 2D image, which is connected with 3D mesh reconstruction part based on GCN via FC layers. The results on 609 experiments showed that the proposed network could achieve higher accuracy in 3D mesh than PLSR-based 3D shape instantiation in \cite{zhou2018real}, while it is slightly lower than KPLSR-based 3D shape instantiation in \cite{zhou2018real}.
According to the result, one-stage shape instantiation directly from 2D image to 3D mesh can be achieved by the proposed Instantiation-Net, obtaining comparable performance with the two baseline methods. 

We believe that the combination of DCNN and GCN will be very useful in the medical area, as it bridges the gap between the image and mesh modality. In the future, we will work on extending the proposed Instantiation-Net to broader applications, i.e. reconstructing 3D meshes directly from 3D volumes.

\section{Acknowledge}
Thank Dr Su-Lin Lee and Qing-Biao Li for the helpful instruction on building the 2D and 3D SSM, and understanding GCN respectively. We gratefully acknowledge the support of NVIDIA Corporation with the donation of the Titan Xp GPU used for this research.

\bibliographystyle{IEEEtran}
\bibliography{ICRA2020.bib}

\end{document}